\documentclass[review]{elsarticle}

\usepackage{lineno,hyperref}
\usepackage{subfigure}

\modulolinenumbers[5]

\journal{Journal of \LaTeX\ Templates}

\begin{document}

\begin{frontmatter}

\title{Explosive, continuous and frustrated synchronization transition in spiking Hodgkin-Huxley neuronal networks: the role of topology and synaptic interaction}% Force line breaks with}

\author{Mahsa Khoshkhou}
\author{Afshin Montakhab \fnref{Montakhab@shirazu.ac.ir}}
\address{Department of Physics, College of Sciences, Shiraz Universiy, Shiraz 71946-84795, Iran}

\begin{abstract}
Synchronization is an important collective phenomenon in
interacting oscillatory agents. Many functional features of the
brain are related to synchronization of neurons. The type of
synchronization transition that may occur (explosive vs.
continuous) has been the focus of intense attention in recent
years, mostly in the context of phase oscillator models for which
collective behavior is independent of the mean-value of natural
frequency. However, synchronization properties of
biologically-motivated neural models depend on the firing
frequencies. In this study we report a systematic study of
gamma-band synchronization in spiking Hodgkin-Huxley neurons which
interact via electrical or chemical synapses. We use various
network models in order to define the connectivity matrix. We find
that the underlying mechanisms and types of synchronization
transitions in gamma-band differs from beta-band. In gamma-band,
network regularity suppresses transition while randomness promotes
a continuous transition. Heterogeneity in the underlying topology
does not lead to any change in the order of transition, however,
correlation between number of synapses and frequency of a neuron
will lead to explosive synchronization in heterogenous networks
with electrical synapses. Furthermore, small-world networks
modeling a fine balance between clustering and randomness (as in
the cortex), lead to explosive synchronization with electrical
synapses, but a smooth transition in the case of chemical
synapses. We also find that hierarchical modular networks, such as
the connectome, lead to frustrated transitions. We explain our
results based on various properties of the network, paying
particular attention to the competition between clustering and
long-range synapses.
\end{abstract}

\begin{keyword}
\texttt{ Synchronization, Hodgkin-Huxley neuron, phase transition, electrical and chemical synapses, complex networks}
%\MSC[2010] 00-01\sep  99-00
\end{keyword}

\end{frontmatter}

%\linenumbers

\section{Introduction}
The phenomenon of phase transition is an important part of modern
statistical physics with many applications in physical and
biological systems
\cite{Callen1998,Ma2018,Weisbuch2018,Steyn-Ross2010}. It is
generally believed that many naturally occurring systems
self-organize to the edge of a phase transition point which can
lead to many functional advantages
\cite{Markovic2014,Bak1987,Bak1989}. In the case of biological
systems, such transitions are generally believed to be of the
critical (continuous) nature associated with a critical point
\cite{munoz2018,dante2010} or an extended critical
regime\cite{MM2013,MMV2017}. Synchronization transition is an
interesting phase transition that might occur in some important
systems such as power grids \cite{Rohden2012}, ecological systems
\cite{Blasius1999}, and seasonal epidemics spreading
\cite{Stone2007}, with both a continuous as well as a more
interesting explosive transition.  It is also interesting to note
that the critical brain hypothesis \cite{dante2010} had originally
assumed that the brain operates at the edge of an \emph{activity}
phase transition. However, recent theoretical
\cite{diSanto2018,KM2019} as well as experimental \cite{fonte2019}
studies show that the criticality may be associated with a
\emph{synchronization} transition. This possibility provides a
stronger motivation to study various types of synchronization
transition that may occur in network models of biological neurons.

Phase synchronization also has a key role in large-scale
integration, memory, vision and other cognitive tasks performed by
the human brain \cite{Varela,Buehlmann2010,Esir2017,Fell}. In a
healthy brain, synchrony must occur at a moderate level. Excessive
synchronization leads to brain disorders like epilepsy or
Parkinson, while schizophrenia and autism are related to deficit
of synchronization among neurons \cite{Kandel}. Thus a healthy
brain is thought to be functioning at the edge of synchronization
transition between order and randomness \cite{Gireesh2008,KM2019,diSanto2018}. From this perspective, a
slight increase in neural interactions might lead to a
synchronization transition in local neural circuits. The type of
resultant transition (continuous, explosive or frustrated) is
therefore important. For example, when the emerging transition is
a continuous one, then a small change in the interaction strength
changes the amount of synchronization slightly. But if the
emerging transition is an explosive one a small increase in the
interaction strength may result in a sudden emergence of global
order in the neural circuit. Explosive synchronization has
functional advantages if it occurs during a fast response, but it
also has disadvantages if it occurs, for example, during an
epileptic seizure.

Brain oscillations are categorized in various frequency bands. For
example, beta-band (13-30 Hz) are typically associated with
cognitive task performance. Recently, we have provided a
systematic study of beta-band synchronization transitions in
network models of Izhikevich neurons \cite{KM2018} and showed that
contrary to the case of simple phase oscillators, biologically
meaningful models of neural dynamics exhibit synchronization
transitions which depend on the average firing frequency of
neurons \cite{KM2018}. This difference is rooted in the fact that
phase oscillator dynamics has a single time-scale (the mean-value
of natural frequencies) which can be re-scaled without having any
significant influence on the dynamics of the network \cite{Gros},
while biologically plausible neural dynamics typically has more
than one time-scale, e.g. refractory period.  The
frequency-dependent behavior can arise when one of these
time-scales depends on a changing parameter, while the other one
does not, thus leading to changing ratio of the various
time-scales \cite{KM2018}. In fact it was shown that the patterns
of transition changed significantly when one increased the average
frequency to the gamma-band ( $>$30 Hz).

Gamma-band oscillations are also an important class of rhythms
appearing during a broad range of the brain activities
\cite{Buzsaki}, and have received a great deal of attention.
Gamma-band oscillations have been observed in several cortical
areas, as well as subcortical structures \cite{Jia2011}. In
sensory cortex, gamma power increases with sensory drive
\cite{Henrie2005}, cognitive tasks including feature
binding\cite{Fries2007}, visual grouping \cite{Vidal2006},
stimulus selection \cite{Berns2008,Liu2006} and attention
\cite{Fries2001}. In higher cortex, gamma power is the dominant
rhythm during working memory \cite{Roux2014} and learning
\cite{Bauser2007}. Also, it is reported that irregular gamma waves
have been observed in pathologies such as Alzheimer
\cite{Uhihaas2006}.

Our purpose here is to provide a systematic study of a
biologically motivated neuronal network. We therefore propose to
study synchronization transition in gamma-band and seek the effect
of synaptic interaction (chemical vs. electrical synapses) as well
as the topology of the network used on the ensuing transition
type. However, Izhikevich neurons have a tendency to burst as
opposed to spike when one increases the input in order to increase
the frequency. Furthermore, increasing interaction strength in
network of Izhikevich neurons also leads to bursting behavior.  On
the other hand Hodgkin-Huxley (HH) neurons have a large stable
spiking range in gamma frequencies \cite{Hodgkin}. We therefore
use network models of HH neurons in gamma band in order to study
synchronization patterns which emerge.

Although synchronization of HH (or HH-type) neurons has been
extensively studied before, e.g. in
\cite{Perez2011,Wang2000,Kwon2002,Kwon2011,Wang2008,Park2011,Zhou2003,Prado2014,Batista2013},
a systematic study of (the order of) synchronization transition
has not been performed to the best of our knowledge. In fact, much
of such type of studies usually employ phase oscillator models
such as the Kuramoto model \cite{Kuramoto1987,Acebron2005}. Here,
our emphasis is to ascertain the type of phase transition (e.g.
continuous vs. explosive) that may occur in a collection of HH
neurons \textit{and} how that may depend on synaptic interaction
and/or underlying structure (network) \cite{MM2015}. Surprisingly,
we find that one- and two-dimensional lattice networks of spiking
HH neurons exhibit no transition. Instead they exhibit
quasiperiodic partial synchronization as a result of strong
clustering which does not lead to global order due to lack of
long-range interactions. Random network structures like
Erdos-Renyi (ER) and scale-free (SF) networks exhibit continuous
transition with either electrical or chemical synapses, with no
significant difference between SF and ER structures. However,
small-world network with high clustering coefficient and
long-range interaction exhibits explosive (first-order) transition
to synchronization when neurons interact via electrical synapses,
but exhibits continuous (second-order) transition when interacting
via chemical synapses. Furthermore, we consider the role of
heterogeneity by introducing a correlation between frequency and
the degree of a given neuron. We find that while heterogeneity (in
degree or frequency) does not change the order of continuous
transition, a correlation between the two can lead to explosive
synchronization with electrical synapses, but not with chemical
synapses. Finally, we show that hierarchical modular (HM) networks
with both types of synapses exhibit frustrated synchronization in
an intermediate regime between disordered and ordered phases of
the system. Some of the structures studied here have been studied
in the beta-band and will consequently be compared and contrasted.
However, the case of correlated heterogeneity as well as HM
networks are just included in the current study and their
counterparts in beta-band had not been studied in
ref.\cite{KM2018}. Consequently, such results can be compared with
those of phase oscillators independent of frequency.

In the following section, we describe the model we use for our
study. In Section (3), we describe our simulation details
including the numerical methods used. Extensive results of our
numerical study are presented in Section (4), and we close the
paper with some concluding remarks in Section (5).

\section{Model}
Consider $N$ Hodgkin-Huxley neurons on an arbitrary network. Electrical activity of $i^{th}$ neuron of the network is described by a set of four nonlinear coupled ordinary differential equations as follows\cite{Hodgkin}:
\begin{eqnarray}\label{eq1}
 &C_m \frac{dv_i}{dt}=I_i^{DC}+I_i^{syn}-G_{Na}m_i^3h_i(v_i-V_{Na}) \nonumber \\
&-G_Kn_i^4(v_i-V_K)-G_L(v_i-V_L)
\end{eqnarray}
\begin{equation}\label{eq2}
\frac{dm_i}{dt}=\alpha_m(v_i)(1-m_i)-\beta_m(v_i)m_i
\end{equation}
\begin{equation}\label{eq3}
 \frac{dh_i}{dt}=\alpha_h(v_i)(1-h_i)-\beta_h(v_i)h_i
\end{equation}
\begin{equation}\label{eq4}
\frac{dn_i}{dt}=\alpha_n(v_i)(1-n_i)-\beta_n(v_i)n_i
\end{equation}
\\
for $i=1, 2, ..., N$. Here $v_i$ is the membrane potential, $m_i$ and $h_i$ are variables for activation and inactivation of sodium current, and $n_i$ is the variable for activation of potassium current \cite{Hodgkin}. $\alpha$ and $\beta$ functions are the so-called rate variables of HH neuron for each type of ionic currents and depend on the instantaneous membrane potential \cite{Hodgkin}. We use $C_m=1.0$, $G_{Na}=120$, $G_K=36$, $G_L=0.3$, $V_{Na}=50$, $V_K=-77$ and $V_L=-54.387$ for the constant parameters \cite{Perez2011}. $I_i^{DC}$ is an external current which differs from a neuron to the other and determines dynamical properties of uncoupled HH neurons. It is shown that for $I^{DC}>9.8 {\mu}A/cm^2$ a stable limit-cycle is the global attractor for a single HH neuron \cite{SGLee}. We choose values of $I_i^{DC}$ randomly from a Poisson distribution with mean value $10.0 {\mu}A/cm^2$. Therefore, intrinsic firing rates are non-identical and most of the neurons spike regularly with gamma rhythms \cite{Buzsaki}. Here, we set the mean intrinsic firing rate is $f{\simeq}75$ Hz, unless otherwise stated.

The term $I_i^{syn}$ in Eq.(1) represents synaptic current received by post-synaptic neuron $i$. Functional form of this current depends on the synaptic type. For a gap junction or an electrical synapse the synaptic current is \cite{Roth}:
\begin{equation}\label{equ5}
I_i^{syn}=\frac{1}{D_i}\sum_jg_{ji}(v_j-v_i)
\end{equation}
\\
and if the synapse is chemical then \cite{Roth}:
\begin{equation}\label{equ6}
I_i^{syn}=\frac{1}{D_i}\sum_jg_{ji}\frac{exp(-\frac{t-t_j}{\tau_s})-exp(-\frac{t-t_j}{\tau_f})}{\tau_s-\tau_f}(V_0-v_i)
\end{equation}
\\
where $D_i$ is in-degree of node $i$, $g_{ji}$ is the strength of synapse from pre-synaptic neuron $j$ to post-synaptic neuron $i$. Here we assumed that all existing synapses have the same strength, viz $g_{ji}=g a_{ji}$, where $g$ is the electrical conductance of synapse and $a_{ji}$ is the element of adjacency matrix of the underlying network. Also in Eq.(6) $t_j$ is the instance of last spike of pre-synaptic neuron $j$, $\tau_s$ and $\tau_f$ are the slow and fast synaptic decay constants and $V_0$ is the reversal potential of synapse which is equal to zero since we assumed that all synapses in our circuit are excitatory. In this study we take $\tau_s=1.7$ and $\tau_f=0.2$ which are the values obtained according to experimental data \cite{Roth}.  From the functional form of these synaptic currents, one can expect that they might have different effects on synchronization of neuronal networks. For example, electrical synapses depend on the phase difference of connected neurons with increasing strength for unsynchronized neurons, while chemical synapses tend to effect post-synaptic neurons regardless of the phase difference, and decaying in strength as a function of time after pre-synaptic firing time $t_j$. Therefore, for example, one would expect for a given value of coupling strength $g$, electrical synapses would provide more synchronization when compared to chemical synapses.

In order to quantify the amount of phase synchronization in a neural population, we assign an instantaneous phase to each neuron as in \cite{Pikovsky1997}:
\begin{equation}\label{equ7}
\phi_i(t)=2\pi\frac{t-t_i^m}{t_i^{m+1}-t_i^m}
\end{equation}
\\
where $t_i^m$ is the instant of $m^{th}$ spike of neuron $i$. Then we define a global instantaneous order parameter as:
 \begin{equation}\label{equ8}
 S(t)=\frac{2}{N(N-1)}\sum_{i\neq j}cos^2\Big{(}\frac{\phi_i(t)-\phi_j(t)}{2}\Big{)}
 \end{equation}
\\
The global order parameter $S$ is the long-time-average of $S(t)$
at the stationary state and measures collective phase
synchronization in oscillations of membrane potentials  of all
neurons, viz $S={\langle}S(t){\rangle}_t$. $S$ is bounded between
0.5 and 1. If neurons spike out-of-phase, then $S{\simeq}0.5$
where they spike completely in-phase $S{\simeq}1$. For states with
partial synchrony $0.5<S<1$.

Along with the order parameter $S$, we have also calculated the
more commonly used Kuramoto order parameter \cite{Kuramoto1987}:
\begin{equation}\label{equ9}
R(t)e^{i\theta}=\frac{1}{N}\sum_je^{i\phi_j(t)}
\end{equation}
\\
with $R=\langle R(t) \rangle_t$ where $0 \leq R \leq 1$. $R=0$
indicates asynchronous, while $R=1$ completely synchronous,
oscillations. Essentially, the same results are obtained for $R$
as those obtained for $S$. However, from a statistical point of
view $R(t)$ represents an average of $N$ data points while $S(t)$
represents an average of $N(N-1)/2$ data points which results in
better statistics for our limited system sizes, and therefore
better statistics considering our system size limitations. We also
define a generalized susceptibility as the relative
root-mean-square fluctuations in the given order parameter:
\begin{equation}\label{equ10}
 \kappa_S=\Big{(}\frac{\langle S^2(t)\rangle - \langle S(t) \rangle^2}{\langle S(t) \rangle^2}\Big{)}^{1/2}
\end{equation}
\\ or:
\begin{equation}\label{equ11}
\kappa_R=\Big{(}\frac{\langle R^2(t)\rangle - \langle R(t) \rangle^2}{\langle R(t) \rangle^2}\Big{)}^{1/2}
\end{equation}
\\ Such generalized susceptibilities are very useful tools in order
to study phase transitions in general, since critical systems are
supposed to exhibit maximal fluctuations at the critical point,
diverging in the thermodynamic $N\rightarrow \infty$ limit.

\section{Methods}
We have scrutinized transition to phase synchronization in
networks with $N$ HH neurons interacting via two different
synaptic types. We start by providing a detailed description of
our procedure. We first determine the network topology by
specifying elements of its adjacency matrix. These elements are
either zero or one depending on if the nodes are unconnected or
connected, respectively. The links in our networks are symmetric.
Synapses are also not plastic in this study. The strength of
synapses is set with parameter $g$ explained in the previous
section. After constructing each network, the synaptic type is
determined. If synapses are supposed to be electrical, we use
Eq.(5) to describe synaptic currents. While neurons are assumed to
interact via chemical synapses, Eq.(6) is used. Next, we fix the
values of $I_i^{DC}$ and set the parameter $g$ equal to zero. We
then integrate Eqs.(1)-(4) using fourth order Runge-Kutta method
with a fixed time step ${\Delta}t=10^{-3}$ ms. Typically, much
larger time steps are used in simulations of HH neurons, see for
example \cite{Kwon2002,Kwon2011}.  However, since long relaxation
times were required in our studies (particularly near the
transition points) we choose such a short time step in order to
avoid the accumulation of errors. Using this small time step, we
are able to specify $t_i^m$, the instant of $m^{th}$ spiking of
each neuron with an accuracy of $10^{-3}$ ms. Finally we obtain
the phase of all neurons and calculate $S(t)$ and $R(t)$ at every
time instant (Eqs.(8) and (9)). We allow the dynamics to progress
for a long transient time (order of $10^{6}$ time steps) until the
fluctuations in $S(t)$ or $R(t)$ reach a stationary state. After
reaching stationary state, we run our simulation for another
$2{\times}10^{4}$ ms ($2{\times}10^7$ time steps) and evaluate the
order parameter $S$ and $R$ by averaging $S(t)$ and $R(t)$ over
this second interval. We next increase the value of $g$ slightly
(keeping all other conditions fixed) and repeat the whole process
to evaluate $S$ and $R$ again. In this manner we obtain dependence
of order parameter on coupling strength $g$, in each network
topology and for each of the above mentioned synaptic types. The
initial condition of integration are random for $g=0$ and the
system is evolved quasi-statically for larger $g$ values. The
synchronization diagrams that are reported here are results of
averaging over five network realizations as well as other
stochastic parameters. Our results are reported for typically
$N\approx500$, but the limited system size does not seem to be an
issue in the results to be presented, as essentially the exact
same results was obtained when we changed the system size within
the range of our computational limits.

\section{Results}
\subsection{Regular networks}
\begin{figure*}[!htbp]
\begin{center}
{\includegraphics[width=1.0\textwidth,height=0.75\textwidth]{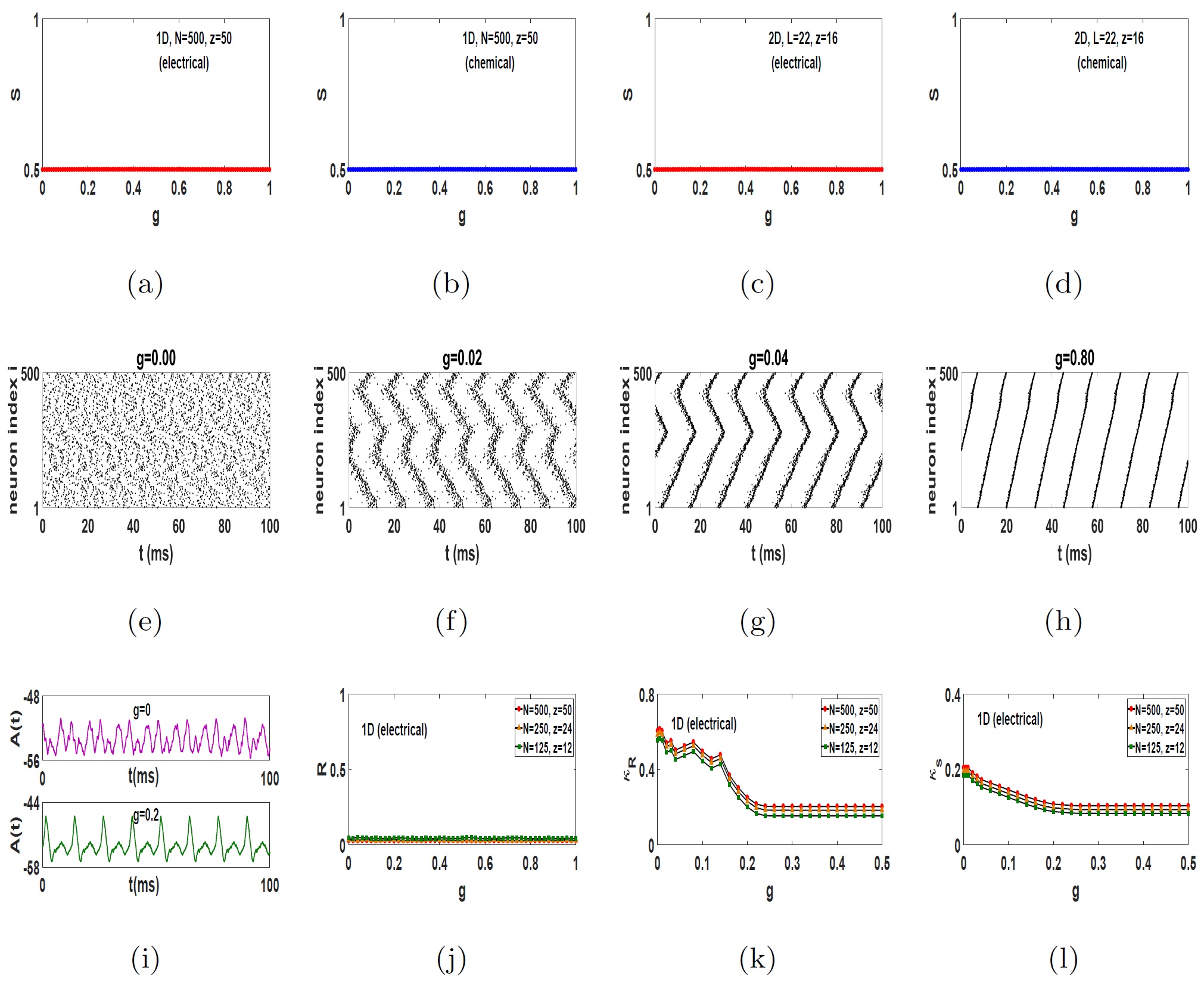}}
\end{center}
\caption{\small Synchronization diagram of HH neurons on regular networks: (a) and (b) One-dimensional ring with electrical and chemical synapses. (c) and (d) Two-dimensional lattice with electrical and chemical synapses. (e)-(h) Raster plots of the one-dimensional ring with electrical synapses for four values of $g$. (i) Network activity for the system in fully asynchronized (orange curve) and in quasiperiodic partially synchronized (green curve) states. (j) The Kuramoto order parameter $R$ vs $g$ for 1D rings with electrical synapses for different system sizes $N$. (k) and (l) The generalized susceptibilities $\kappa_R$ and $\kappa_S$ vs $g$ for the same systems as in (j). $z=0.1N$ in each case. $t=0$ indicates the beginning of stationary state. The synchronization diagrams and susceptibility plots show the averaged results over five initial conditions.} \label{fig1}
\end{figure*}

The first structure that we consider is regular network, a one
dimensional ring of size $N=500$ and $z=50$ as well as a
two-dimensional lattice of side $L$ ($N=L\times L$) with $L=22$
and $z=16$. $z$ is the coordination number of the network. The
results are shown in Figs.1(a-d). It is observed that increasing
$g$ does not lead to a transition in either case. It is somewhat
surprising as one would expect a transition to synchrony for large
$g$. We have therefore investigated the raster plots for this
system for different $g$ values. Such raster plots for
one-dimensional ring with electrical synapses for four values of
$g$ are shown in Figs.1(e-h). Raster plots for rings with chemical
synapses are qualitatively the same as Figs.1(e-h). It is realized
that imposing a small interaction among neurons in a regular ring
leads to formation of correlated regions. This is not unexpected
since regular rings have high clustering coefficient \cite{Gros}.
Increasing $g$ slightly, regulates the phase of neurons on a local
level. Since there are no long-range synapses in the system,
further increase of $g$ could not vanish phase lags among neurons
belonging to far away areas of the network, but instead results in
the emergence of the so-called quasiperiodic partial
synchronization. Quasiperiodic partial synchronization is denoted
to the state of a population of interacting oscillators in which
the system sets into a nontrivial dynamical regime where
oscillators display quasiperiodic dynamics while collective
observable of the system oscillate periodically
\cite{vanVreeswijk,Rosenblum2007,Burioni}. A relevant collective
observable for a neural network is the network activity that is
defined as $A(t)=\frac{1}{N}\sum_{i=1}^Nv_i(t)$. In Fig.1(i) we
have plotted A(t) for a regular ring of HH neurons for $g=0.00$
when neurons are fully asynchronous (orange curve) and also for
$g=0.80$ which is where the network is in a quasiperiodic partial
synchronization state (green curve). It is seen that when neurons
spike out of order, $A(t)$ fluctuates irregularly. But when the
network is in state of quasiperiodic partial synchronization
neurons spike quasiperiodically and $A(t)$ oscillates
periodically. This state emerges in the rings from $g{\simeq}0.20$
and remains robust when $g$ is increased further as the
perspective of raster plots remain qualitatively the same from
$g=0.20$ to $g=1.00$ (or even for larger values of $g$ which are
not shown here).

For sake of comparison, in Fig.1(j) we show the $R-g$ plot for 1D
ring with electrical synapses for three different system sizes
$N$. The coordination number in each system is set to be $z=0.1N$.
In light of these plots we find that the synchronization diagram
remains unaltered upon increasing the system size. Also, comparing
Figs. 1(a) and 1(j) one verifies the equivalence of the results
obtained based on the order parameters $R$ and $S$, except for the
more refined statistics resulting from $S$. Moreover, the
generalized susceptibilities $\kappa_R$ and $\kappa_S$ for the
same systems as in Fig.1(j), are illustrated in Figs. 1(k) and
1(l), respectively. It is observed that increasing $g$ does not
lead to any distinctive peak in $\kappa_R$ or $\kappa_s$,
confirming that no phase transition occurs in this systems.
Generalized susceptibilities for other regular networks studied
here are qualitatively similar to Figs.1(j) and 1(l) (not shown).

We note that one might suspect that the lack of transition
observed in the 1D lattice might be due to the low dimensional
structure, similar to the lack of phase transition in, for
example, 1D Ising model. That is why we have also performed
simulations for the 2D $L \times L$ lattice whose main results are
shown in Fig.1(c) and 1(d) which again show no transition.  We
note that raster plots of the 2D system are quantitatively the
same as the 1D case with lesser coherence (due to smaller
clustering) and that network oscillations, $A(t)$, are also very
similar to the 1D case (not shown).

\subsection{ER and SF networks}
\begin{figure*}[!htbp]
\begin{center}
{\includegraphics[width=1.0\textwidth,height=0.75\textwidth]{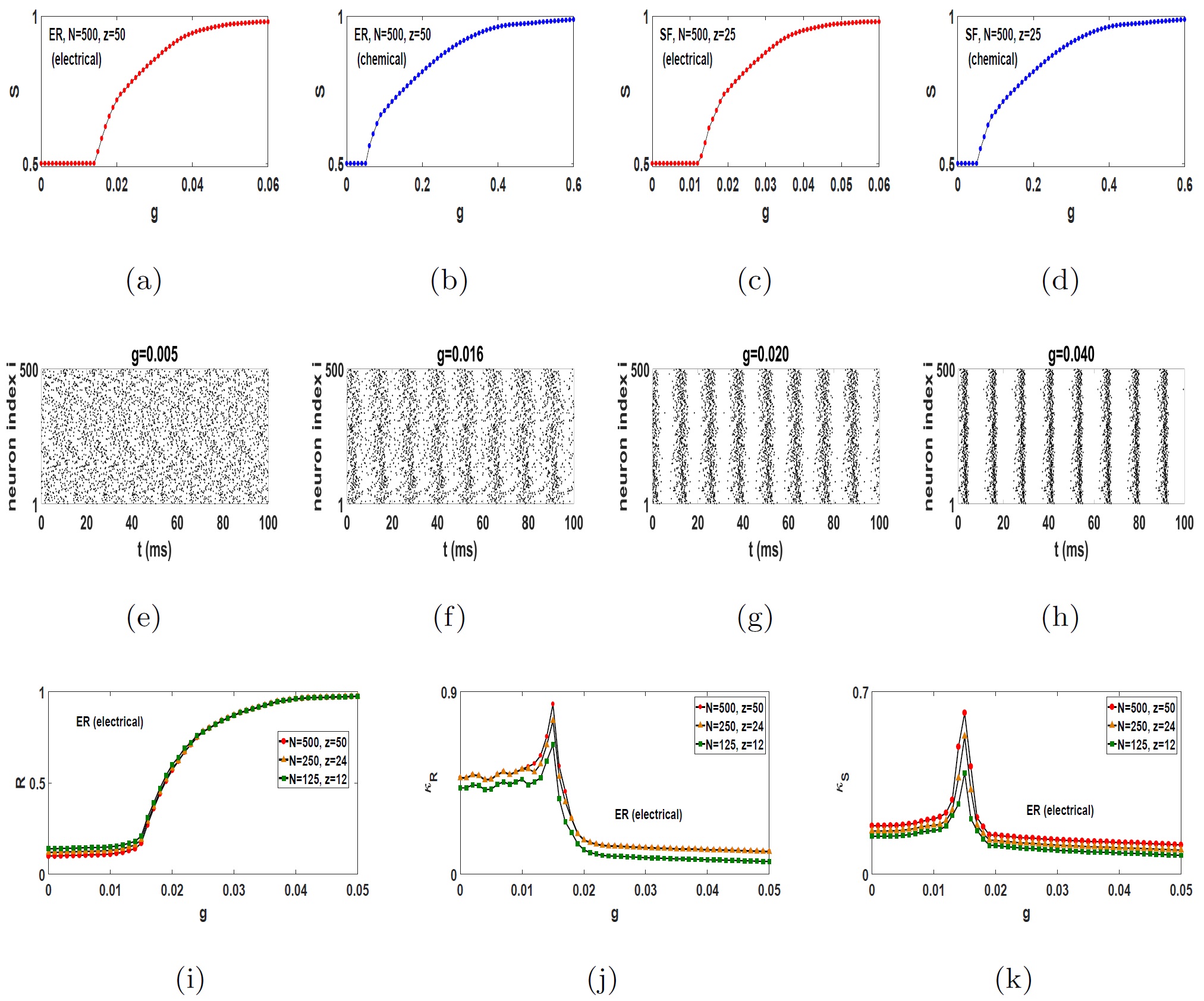}}
\end{center}
\caption{\small (a) and (b)Transition to phase synchronization in ER networks of spiking HH neurons with electrical and chemical synapses. (c) and (d) Transition to phase synchronization in SF networks of spiking HH neurons with electrical and chemical synapses. (e)-(h) Raster plots for ER network with electrical synapses for various values of $g$. (i) Kuramoto order parameter $R$ vs $g$ for ER networks of HH neurons with electrical synapses for different system sizes $N$. (j) and (k) $\kappa_R$ and $\kappa_S$ vs $g$ for the same systems as in (i). $z=0.1N$ in each case. $t=0$ indicate the beginning of stationary state. The synchronization diagrams and susceptibility plots show the averaged results over five network realizations and initial conditions.}
\label{fig2}
\end{figure*}

We next consider random networks with small-world effect but with
much smaller clustering compared to regular networks. We consider
ER network which has a homogeneous random structure as well as SF network
which has a heterogeneous random structure \cite{Gros}. Such
networks are constructed using a configurational model
\cite{Gros}. The networks size is $N=500$. $z=50$ for ER network
and $z=25$ for SF network. Also the degree distribution function
of SF network is $P(k){\sim}k^{-\gamma}$ with $\gamma=2.2$. The
results for synchronization transition of HH neurons with
electrical and chemical synaptic currents, on ER and SF are shown
in Figs.2(a-d). It is observed that the system with both types of
synapses exhibits a continuous transition from asynchrony to
synchrony. Raster plots of spikes for the ER network with
electrical synapses are illustrated in Figs.2(e-h). It is evident
that since clustering coefficient is significantly reduced due to
randomness (as compared to regular networks), neuronal clusters do
not appear in the system. However, presence of a significant
number of long-range connections regulates neural activity in this
random network when $g$ is increased above a certain threshold.
Raster of spikes for other transitions are qualitatively similar
to those of Figs.2(e-h) (not shown). Looking at the value where the
transition occurs, $g_t$, for a given network, one concludes that
synchronization is more conducive to electrical synapses than
chemical synapses, i.e. $g_t$ is about an order of magnitude
smaller for electrical synapses. This makes sense as electrical
synapses are known to be stronger than chemical synapses. On the
other hand, the strong similarity between the results for ER and
SF networks for a given synaptic type, including their
corresponding value at transition, leads one to conclude that the
role of structural heterogeneity (SF network) is not an important
factor in influencing the type and shape of transition curves
($S-g$ plots).

In Fig.2(i), we also show $R-g$ plots of HH neurons with
electrical synapses on ER networks with three system sizes $N$ to
be compared with with the $S-g$ plot in Fig.2(a). Here, $z=0.1N$
in each system size. Furthermore, variations of the generalized
susceptibilities $\kappa_R$ and $\kappa_S$ upon increasing $g$ for
the systems of Fig.2(i) are plotted in Figs.2(j) and 2(k),
respectively. It is observed that both $\kappa_R$ and $\kappa_S$
show a specific peak at the transition point which grows with
increasing the system size. The behavior of the generalized
susceptibilities further collaborates our order parameter results
which indicate that our model does not show synchronization
transition for low dimensional systems (Fig.1), but exhibits
definitive and continuous transition in a high dimensional
structure such as complex networks (Fig.2).

Regarding the results associated with figures 1 and 2,  we can
conclude two important points: (i) the main results are unaltered
upon increasing the system size $N$, and (ii) the synchronization
diagrams exhibit qualitatively the same behavior whether we employ
$R$ or $S$, except for the more refined statistics provided by $S$
which enables us to determine the transition point clearly.
Therefore, for the rest of this paper we report the results only
based on the order parameter $S$ and for our largest available
system size ($N \simeq 500$).

\subsection{Small-world networks}
After considering regular networks with high clustering but large
average distance on one hand, and highly random networks with
strong small-world effect but negligible clustering on the other
hand, we are interested in networks that have high clustering
coefficient, as well as small-world property. Therefore, we
constructed Watts-Strogatz (WS) networks \cite{Gros} with $N=500$
and $z=50$ by random rewiring of two percent of links of a regular
ring. This low rewiring probability ($p=0.02$) allows the system
to keep its large clustering coefficient while developing
significantly low average distance (i.e. small-world effect).  The
resulting $S-g$ curves for WS networks with electrical and
chemical synapses are shown in Fig.3(a) and 3(b), respectively.
Interestingly, we observe a discontinuous (explosive) transition
for the case of electrical synapses while a continuous transition
is observed for the chemical synapses. As we will discuss shortly,
this type of explosive synchronization is different in its
mechanism than those seen for phase oscillators in heterogeneous
networks such as \cite{Gardenes2011,Leyva}. The synchronization
transition is accompanied with a hysteresis loop if a backward
sweep in $g$ is performed from the highly synchronized state.
Therefore, as seen in Fig.3(a), not only the transition is
explosive, the value of $S$ is also history-dependent. This is to
be contrasted with the case of chemical synapses in WS network
where increasing $g$ leads to a continuous transition from
asynchrony to synchrony in neural spiking as is clear in Fig.3(b).

\begin{figure}[!htbp]
\begin{center}
{\includegraphics[width=1.0\textwidth,height=0.75\textwidth]{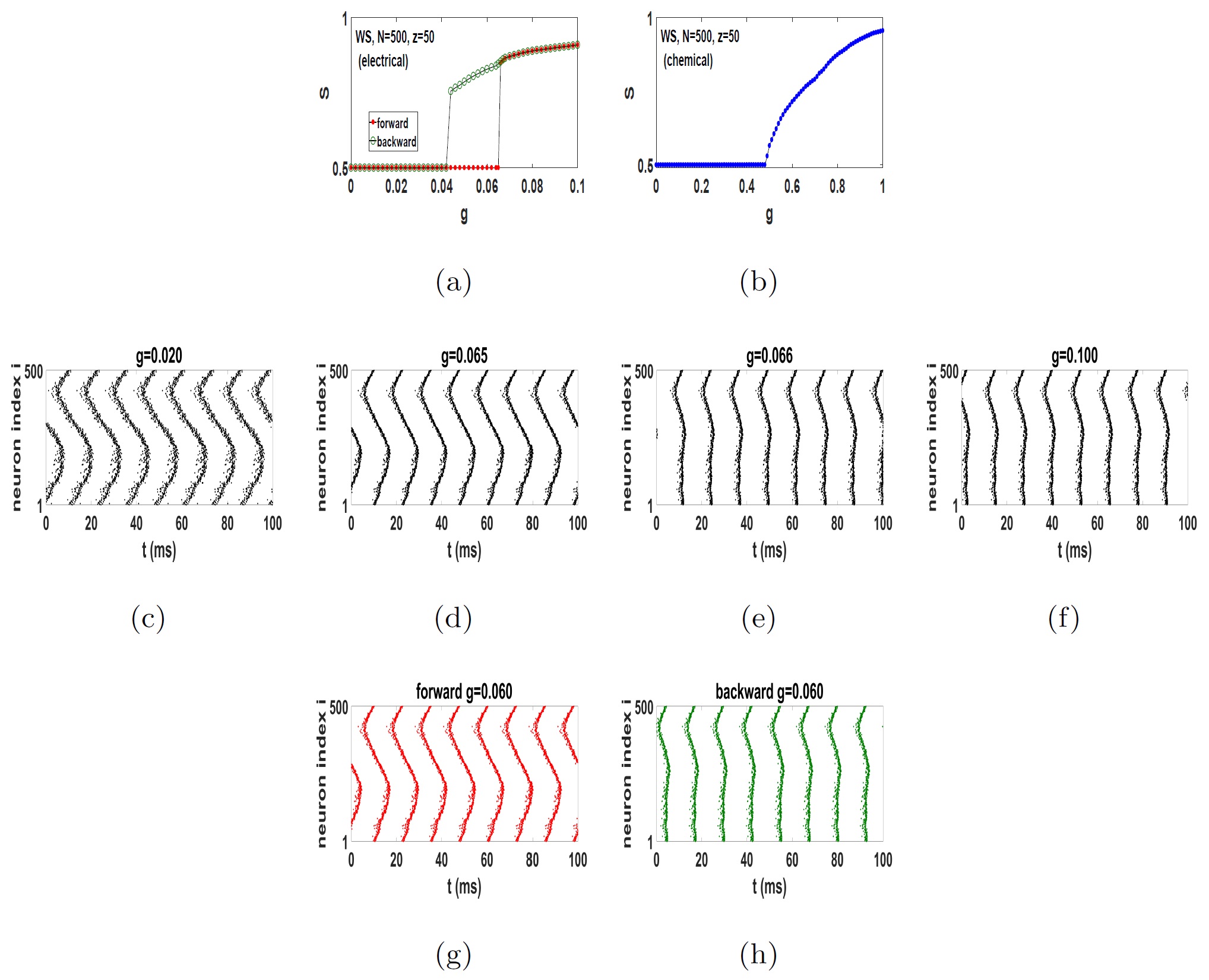}}
\end{center}
\caption{\small Transition to phase synchronization in WS networks of spiking HH neurons with (a) electrical and (b) chemical synapses. Network size and coordination number are $N=500$ and $z=50$, respectively and rewiring probability is $p=0.02$. (c)-(f) Raster plots for the network with electrical synapses for various values of $g$ of the system in forward direction. (g) and (h) Two raster plots for the same value of $g$ inside the hysteresis loop ($g=0.060$) in forward and backward evolution of the system. $t=0$ indicate the beginning of stationary state. The synchronization diagrams show the averaged results over five network realizations and initial conditions.} \label{fig3}
\end{figure}

Therefore, one- and two-dimensional regular networks produced no
transition, while random networks produced a continuous
transition. However, small-world networks which lie somewhere
between randomness and regularity exhibit explosive
synchronization (electrical) as well as continuous transition
(chemical). The fact that transition type in WS network depends on
the interaction type is an interesting result and may be important
from the point of view of neuroscience, since it has been reported
that the brain networks at the microscopic level are similar to WS
networks \cite{Sporns2011}. To elucidate the effect of topology
and underlying reason for different order of phase transitions, we
display the raster plots of HH neurons with electrical synapses on
a WS network in Fig.3(c-f) for the evolution of system in the
forward direction. Here, the combined effect of clustering and
long-range interaction leads to explosive synchronization.  As in
the regular rings case the effect of clustering leads initially to
correlated regions which are nevertheless not perfectly
synchronized for faraway regions of the network.  Note the
similarity in Fig.1(g) and Fig.3(d), both of which lead to $S=0.5$
and no net synchronization. However, as the effect of long-range
links in the case of WS network is important, increasing $g$ will
eventually lead to interactions among various parts of the network
which eventually leads to global order in the system and therefore
a phase transition, which was absent in models without long-range
interaction. But, why do we observe an explosive synchronization
for electrical synapses but a continuous transition for chemical
synapses?  This has to do with the fact that long-range links
provide strong interactions for the case of electrical synapses as
phase difference of faraway regions is considerable, while
providing weak interaction for local interactions which are mostly
synchronized. When nonlocal regions suddenly go in synch due to
strong electrical interactions a sudden jump in order parameter is
observed.  This is shown in Figs.3(d) and 3(e) for the two
consecutive values of $g$ ($g=0.065$ and $g=0.066$), at the edge
of transition as global order suddenly arises in the system. On
the other hand, in the case of chemical synapses, a pre-synaptic
neuron interacts with a post-synaptic neuron in a decaying fashion
thus providing a weaker effect which allows various regions of the
network to slowly synchronize with each other and thus lead to a
smooth continuous phase transition. Therefore, in the case of
chemical synapses phase lags among various clusters vanish
gradually (not shown). In Figs.3(g) and 3(h), we plot raster plots
within the hysteresis loop for the same value of $g$, one for the
forward branch and one for the backward branch. Here, we note how
small changes in global patterns of spikes can lead to significant
change in the value of $S$.

\subsection{Correlated heterogeneity}
It might have been expected that heterogeneity in network
structure as in the SF network in Fig.2(b) would have led to a
different transition pattern when compared to homogenous random
network of ER. We note that the importance of the role of
heterogeneity in neural networks has attracted much attention in
recent literature. An important observation in regards to
synchronization in the Kuramoto model was that structural
heterogeneity was not sufficient to lead to different transitional
pattern, but a \emph{correlation} between the frequency and the
degree of the node was the key element that would lead to
explosive synchronization in SF networks \cite{Gardenes2011}. This
means that the high frequency nodes in a network are also the
highly connected nodes, while the low frequency nodes are sparsely
connected. We note that the range of frequency in spiking HH
neurons is relatively limited. However, one may attempt to make a
heterogeneous distribution even in this limited range. We have
therefore studied a SF network of size $N=500$ with $\gamma=2.2$
and $k_{min}=7$, $k_{max}=47$, $z=15$. We have also produced the
same distribution of frequencies with $f_{min}=70$ Hz and
$f_{max}=110$ Hz and have studied the correlated ($f\propto k$)
and the uncorrelated distribution of such frequencies. The results
are shown in Fig.4. As is seen, the correlated case leads to
explosive synchronization along with hysteresis in the case of
electrical synapses, but a smooth transition in the case of
chemical synapses.  We also show  the same results for the
uncorrelated case which indicates that the explosive
synchronization is in fact due to the correlation between the two
heterogeneous distributions of degree and frequency, in the case
of electrical synapses.

It is interesting to note that the mechanism for explosive
synchronization is very different here than that observed in WS
network for electrical synapses.  There, it was the combined
effect of local order, which is achieved for low synaptic weight
$g$, and long-range order which sets in for large values of
synaptic weight, that leads to sudden order and explosive
synchronization in the system, see raster plots in Fig.3. Here, in
the case of correlated heterogeneity, the system is still
essentially in a completely disordered phase just before the
explosive synchronization occurs.  See $g=0.227$ and $g=0.228$
raster plots in Fig.4 which are just before and after the
explosive synchronization transition point. This indicates that
the system truly goes through a sudden change from disordered to
ordered phase.  The cause of such an explosive synchronization can
easily be understood by looking at the ordered raster plots. One
sees that in the synchronous phase the entire system is
oscillating at the frequency of $f\approx110$ Hz which is exactly
the frequency of the only hub in the system.  This indicates the
essential role of the hub in this explosive synchronization.  The
entire network must adjust with the hub and once this happens an
explosive synchronization occurs.  This mechanism is very much
similar to what happens in the case of the well-know explosive
synchronization in the Kuramoto model \cite{Gardenes2011}.
However, we emphasize that the explosive synchronization we have
observed only occurs for the stronger electrical synapses, and we
did not observe any explosive synchronization for chemical
synapses in the range of parameters studied here. We finally note
that explosive synchronization occurs at much higher values of $g$
when compared to the WS case and also exhibits a smaller
hysteresis loop.

\begin{figure}[!htbp]
\begin{center}
{\includegraphics[width=1.0\textwidth,height=0.50\textwidth]{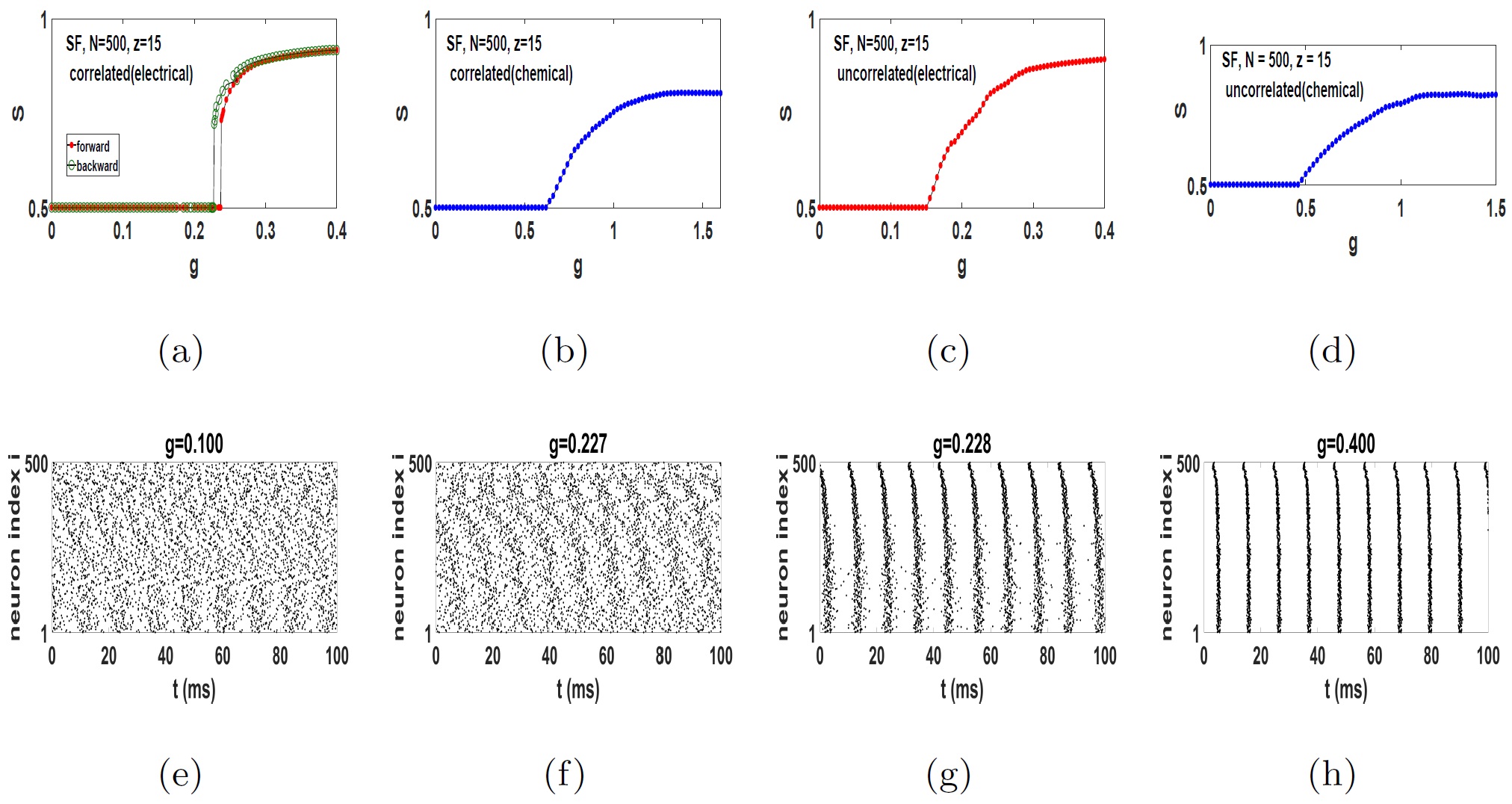}}
\end{center}
\caption{\small Synchronization diagram for the correlated scale-free network (a) electrical synapses and (b) chemical synapses, and the corresponding uncorrelated case (c) and (d). Raster plots for various $g$ are shown in parts (e)-(h) for the case of explosive synchronization in panel (a). $t=0$ indicate the beginning of stationary state. The synchronization diagrams show the averaged results over five network realizations and initial conditions.} \label{fig4}
\end{figure}

\subsection{Hierarchical modular networks}
\begin{figure*}[!htbp]
\begin{center}
{\includegraphics[width=1.0\textwidth,height=0.75\textwidth]{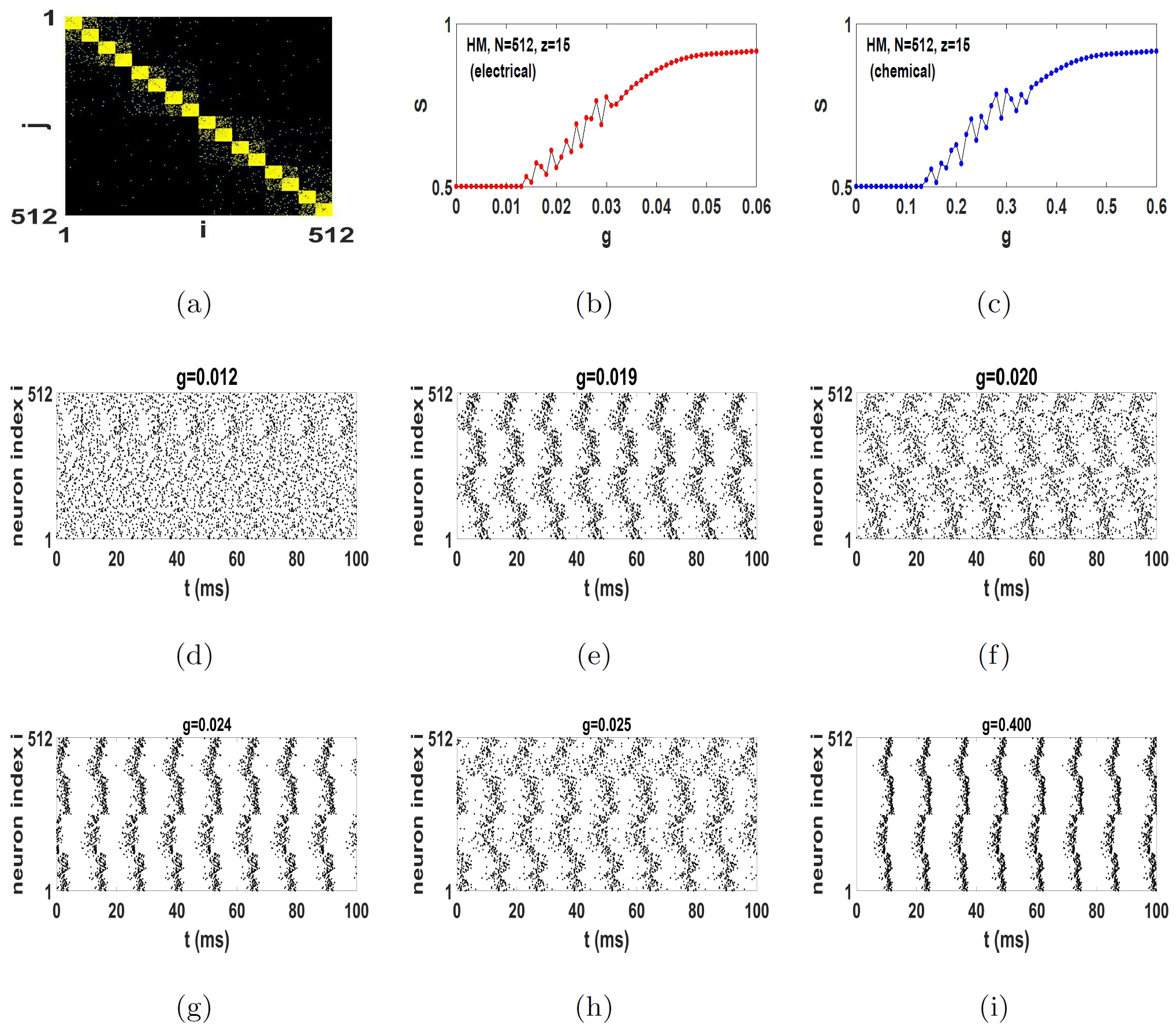}}
\end{center}
\caption{\small (a) Adjacency matrix of the HM network of size $N=512$, coordination number $z=15$. There are 4 hierarchical levels. 16 modules in level 1, 8 modules in level 2, 4 modules in level 3 and 2 modules in level 4. (b) and (c) Synchronization diagram for electrical and chemical synapses, respectively. (d)-(i) Raster plots for neural network with electrical synapses for six different values of $g$. $t=0$ indicates the beginning of stationary state. The synchronization diagrams show the averaged results over five network realizations and initial conditions.} \label{fig5}
\end{figure*}

So far we have investigated synchronization transition of spiking
HH neurons for the most typical network topologies. However, it is
believed that a hierarchical modular (HM) network is a more
realistic representation of the actual structural connectivity of
the cortex on the large scale \cite{Sporns2011}. Therefore it is
worthwhile to investigate synchronization of HH neurons in HM
networks as well. We construct networks with $N=512$ nodes, $z=15$
and 4 hierarchical levels, see Fig.5(a). On the lowest level, the
network is comprised of 16 modules of 32 nodes where each node is
connected to 10 randomly chosen other nodes within the same
module. On the second level, there are 8 modules which are
constructed by connecting pairs of randomly chosen nodes belonging
to two smaller modules of the lower level. In this manner we
construct each module in a higher level by connecting members of
two modules in the previous level. We should note that all links
in the first level are inter-modular and all links in higher
levels are intra-modular. This network despite having considerable
clustering also has small-world effect.

$S-g$ plots for HH neurons with electrical and chemical synapses
for such a HM network is displayed in Figs.5(b) and 5(c),
respectively. We observe that for both types of synapses, there
exists three regimes in the $S-g$ plots. An asynchronous regime
($S=0.5$) for small values of $g$, a synchronous regime for large
values of $g$ and an intermediate regime between ordered and
disordered phases where $S$ does not vary monotonically with
increasing $g$, but reveals a fluctuating behavior. Note that
these fluctuations in order parameter are not due to insufficient
transient time because we have made sure that the system is in its
stationary state before taking measurement of $S$ for each value
of $g$. They also do not appear due to imprecise numerical
integrations as we have taken much care in this regard. Emergence
of this intermediate regime has been reported for the Kuramoto
model on human connectome network which has a hierarchical modular
structure \cite{Villegas}. It is now believed that such
intermediate regime is a manifestation of HM structure of the
underlying network. As can be deduced from the raster plots in
Figs.5(d-i), the HM structure leads to synchronization within
various modules which are themselves out of phase with various
other modules leading to relative synchrony (small $S$) which
nevertheless fluctuates as various modules go in and out of phase
with each other as we change the value of $g$. For example, for
electrical synapses and $g=0.019$ there is more synchronization
than $g=0.020$ as can clearly be seen why from the corresponding
raster plots. Therefore, we observe the same type of
\emph{frustrated} synchronization patterns as in the Kuramoto
model regardless of the synaptic interaction.  We also note that,
looking at the values of transition point $g_t$, one sees strong
similarity with fully random networks of SF and ER (see Fig.2).
This indicates that the onset of synchronization here is also
dictated by long-range links. But, once synchronization sets in,
it is the strong clustering within various modules that dictate
the synchronization pattern for a range of $g$, before global
order sets in for large enough $g$.

\section{Concluding remarks}
In a previous work, we studied beta-band synchronization in
network models of spiking neurons. There, we showed that the type
of synchronization transition occurring in a neural network
depends on the firing rates of constituent neurons \cite{KM2018}.
In this paper we have reported a systematic study of
synchronization transition in network models of spiking neurons in
gamma-band. We employed HH neurons with electrical and chemical
synapses. Our focus has been to characterize the combined effect
of synaptic type and topological features on the type of
synchronization transitions that may occur. The mechanisms and
patterns of synchronization transitions we obtained here for
gamma-rhythms are distinctly different from those we obtained for
beta-rhythms in ref.\cite{KM2018}. For example, in the beta-band
in a one-dimensional lattice, we found a continuous transition for
the electrical synapses, while here for the gamma-band we observed
no transition for any synaptic type in one or two dimensions.
Furthermore, here we found smooth transitions for SF and ER
networks in the gamma-band, while previously we had observed
explosive synchronization on such networks with electrical
synapses. On the other hand, here we observe explosive
synchronization for the WS networks while in the beta-band we only
saw a smooth transition for such networks. We also observed
explosive synchronization in the case of SF networks with
correlated heterogeneity which was not studied in the previous
study for the beta-band.

The underlying mechanisms leading to explosive synchronization in
beta-band was rooted in the formation of anti-phase groups of
neurons for intermediate values of $g$ and their sudden
combination at a transition point \cite{KM2018}. This is
distinctly different from the mechanism that lead to explosive
synchronization in WS network of HH neurons or from the mechanism
resulted in abrupt transition in SF network of HH neurons with
correlated heterogeneity. However, these three mechanisms of
explosive synchronization have a common aspect. They all occur
through electrical synapses. We have not observed explosive
synchronization in the case of chemical synapses. Our results
highlight the fact that electrical synapses are more conducive to
synchronization and can in fact lead to entirely different
transition patterns.  This is in contrary to other studies that
have concluded similar synchronization behavior for electrical and
chemical synapses \cite{Perez2011}.

We also note that it is interesting that our regular one and two
dimensional lattice did not exhibit a transition which is what one
would expect from the study of the Kuramoto model as it, too, does
not exhibit a transition in low dimensional systems \cite{Acebron2005}.  However, the
mechanism for such behavior are different, as we do observe
considerable amount of order in our system with quasiperiodic
oscillations. We once again emphasize the key role of frequency as
well as synaptic interactions in such studies, where in the
beta-band, in one dimension, we had previously observed a
continuous transition for the electrical synapses while no
transition was seen for the chemical synapses.

This brings us to emphasize the difference in the type of
transitions we observed in WS networks. Electrical synapses, which
are strong and fast, lead to explosive synchronization while the
slower and weaker chemical synapses lead to a smooth transition.
This is particulary interesting as neuronal networks are argued to
by on the verge of a phase transition.  This could, for example,
be related to the fact that electrical synapses are useful in fast
involuntary motor response where a strong and fast collective
action is desired, while a smooth transition with chemical
synapses could be understood in terms of cortical neurons where
too much synchronization is deemed to be pathological \cite{Kandel}.

Furthermore, we investigated the role of correlated heterogeneity
and found that in the case of electrical synapses one observes
explosive synchronization while in chemical synapses a smooth
transition occurs.  This result could be interesting from two
aspects. First, it shows that unlike what is generally believed,
correlated heterogeneity does not always lead to explosive
synchronization as chemical synapses showed a smooth transition.
Secondly, it highlights the distinctly different type of explosive
synchronization that may occur in electrical synapses. In WS
network, explosive synchronization occurred \emph{after} the
system gained a high degree of local order, but in the case of
correlated heterogeneity, explosive synchronization was dictated
by the role of the hub with no sign of order in the system just
before the transition occurred.

We have also considered hierarchical modular networks which
resulted in an intermediate regime between order and disorder.
Such a behavior has been previously shown to occur in the Kuramoto
model \cite{Villegas} and our results indicate that such frustrated transition is
a more general property of neuronal systems in HM networks, and it
is furthermore independent of the type of synaptic interaction.

We note that our choice of spiking HH neurons naturally limited
our range of frequency to the gamma band.  However,  we observed
the same type of synchronization patterns when we increased the
natural spiking frequency of the HH neurons to the high gamma band
up to $f \approx 110$, not shown here. We previously observed that
the synchronization patterns changed in the case of Izhikevich
neurons when one increased the average frequency from beta to
gamma band \cite {KM2018}. This was shown to be due to the
dependence of refractory period on the frequency of the neurons.
However, it seems like for high enough frequency the change in the
refractory period becomes negligible due to the short spiking
intervals.

The role of refractory period, conduction/axonal delays, as well
as synaptic plasticity on synchronization patterns are all
potentially interesting avenues for future studies \cite{asl2018}.
Evidence for robust collective oscillations at the edge of chaos,
where scale-invariant activity emerges in neural networks, is
reported in recent experimental and theoretical studies
\cite{Gireesh2008,KM2019,diSanto2018}. Investigation of such
coexistence at the edge of continuous and explosive transitions
obtained in the current study is interesting, as well, although
such investigation will be computationally expensive. One might
also consider the role of external oscillatory input in such
studies which has been recently shown to introduce critical
oscillations in certain models of excitable nodes
\cite{Moosavi2018}. Lastly, the role of noise was absent in our
studies. Noisy dynamics may add some important features to the
collective dynamics including important effects in the nature of
the phase transition \cite{MM2014}.

\section{Acknowledgements}
Support from Shiraz University research council is kindly acknowledged. This work has been supported in part by a grant from the Cognitive Sciences and Technologies Council.

\section*{References}
\bibliography{mybibfile}

\end{document}